\documentclass[12pt]{article}
\def\be{\begin{equation}} \def\ee{\end{equation}} 
\def\bea{\begin{eqnarray}} \def\eea{\end{eqnarray}} 

\makeatletter \def\section{\@startsection {section}{1}{\z@}{-3.5ex 
plus -1ex minus -.2ex}{2.3ex plus .2ex}{\large\bf}} 
\def\subsection{\@startsection{subsection}{2}{\z@}{-3.25ex plus -1ex 
minus -.2ex}{1.5ex plus .2ex}{\normalsize\bf}} 
 
\usepackage{amssymb,amsmath,amsfonts,graphicx}
 
\newcommand{\id}{\hbox{1\kern-.27em l}} 
\newcommand{\sid}{\hbox{\scriptsize1\kern-.27em l}} 

\newcommand{\we}{\kern-.1em\wedge\kern-.1em} 
\newcommand{\scal}{\kern-.13em\cdot\kern-.13em}

\newcommand{\II}{I\kern-.09em I}

\newcommand{\Z}{\mathbb{Z}}

\newcommand{\spa}{\ \ ,\ \ \ \ }

\begin{document}
\begin{titlepage}

\rightline{NORDITA-2003/1 HE} 
 
\vskip 2.5cm \centerline{\LARGE ${\cal N}=1$ super Yang-Mills theories and} 
\vskip 0.5cm 
 \centerline{\LARGE wrapped branes} 
 
\vskip 1.5cm

\centerline{\large P. Merlatti} 
 
\vskip .8cm 
 
\centerline{\sl NORDITA} 
 
\centerline{Blegdamsvej 17, 2100 Copenhagen \O, Denmark} 
 
\centerline{\tt  merlatti@nbi.dk} 
 
\vskip 2cm 
 
\begin{abstract} 
\noindent I consider supergravity solutions of D5 branes wrapped on
supersymmetric 2-cycles and use them to discuss relevant features
of four-dimensional ${\cal N}=1$ super Yang-Mills
theories with gauge group $SU(N)$. In particular, using a gravitational dual 
of the gaugino
condensate, it is shown that is possible to obtain the complete NSVZ $\beta$-function. 
It is also described how different aspects of the gauge theory are nicely 
encoded in this supergravity solution.
\end{abstract}

\vfill
{\it Proceedings of the RTN-Workshop
``The quantum structure of spacetime and the geometric nature of fundamental interactions'',
Leuven, September 2002}
\end{titlepage}

\section{The gauge gravity correspondence and wrapped branes}

Recently, it has become more and more evident that a lot of
relevant information about supersymmetric Yang-Mills (SYM)
theories can be obtained by studying their dual supergravity
backgrounds produced by stacks of D branes. This gauge/gravity
correspondence has been throughly investigated in different cases. 
In this contribution I will mainly concern with ${\cal N}=1$ gauge 
theory in four dimensions. One way to get this amount of supersymmetry is
to consider D branes whose world-volume is partially wrapped on
a supersymmetric cycle inside a Calabi-Yau space.
The unwrapped part of the brane world-volume remains flat and
supports a gauge theory. In order to preserve some
supersymmetry, the normal bundle to the wrapped D branes has to be
partially twisted and, as a consequence of this
twist, some world-volume fields become massive and decouple. This
procedure has been used by Maldacena and Nu\~nez in Ref.~\cite{Maldacena:2000yy} to study the pure
${\cal N}=1$ SYM theory in four dimensions.
This is obtained by considering the world-volume theory of the 
D5-branes at energies where both the higher string modes as well 
as the KK excitations on the 2-cycle decouple. The back-reaction 
of the D-branes deforms the original background. The topology of 
the resulting space is in general very different from the starting 
CY space. In this case, as discussed by Vafa in 
Ref.~\cite{Vafa:2000wi}, one expects the resulting space be a 
deformed CY space, where the 2-cycle has shrunk but a 3-sphere has 
blown-up, rendering a ten-dimensional non-singular solution.
As it is the case for gravity duals of 
confining gauge theories, one cannot obtain an exact duality since 
extra degrees of freedom, not belonging to the gauge theory, cannot be 
decoupled within the supergravity regime. 
The question is now whether one can extract information on the gauge 
theory from this dual 
supergravity background. 

This question has been addressed in many papers  \cite{Maldacena:2000yy,
BM,DiVecchia:2002ks,Loewy:2001pq,Apreda:2001qb,Olesen:2002nh} 
and a 
number of information on the gauge theory have been shown to be predicted 
by the dual supergravity background in a precise and quantitative 
way. In particular, in this contribution I will show how to derive from
the supergravity solution the perturbative running of the gauge coupling with 
the corresponding $\beta$-function, the chiral symmetry anomaly,  the 
phenomenon of gaugino condensation with the corresponding breaking of 
the chiral symmetry to $\Z_2$ in the IR.

\subsection{Ten dimensional supergravity solution}

Let us start by summarizing the explicit form of the MN solution. 
This solution is obtained from a non-singular domain wall solution of 
seven-dimensional gauged supergravity \cite{cv}, parameterized by 
coordinates $(x_0,\ldots,x_3,\rho,\theta_1,\phi_1)$, and uplifting to 
ten dimensions along a 3-sphere 
\cite{Cvetic:2000dm,Chamseddine:1999uy}, parameterized by coordinates 
$(\psi,\theta_2,\phi_2)$. The relevant fields (the metric, the dilaton 
and the RR 3-form the D5-branes magnetically couple to) are 
\begin{eqnarray} 
\label{mnsol} ds^2 &=& e^\Phi dx^2_{1,3} + e^{\Phi} \alpha' g_s N 
\left[ e^{2h} \left( d\theta_1^2 + \sin^2 \theta_1 \,d\phi_1^2 \right) 
+ d\rho^2 + \sum_{a=1}^3\left(\sigma^a - A^a \right)^2 \right] \\ 
\label{mnsol1} e^{2 \Phi} &=& \frac{\sinh 2\rho}{2\, e^{h}} 
\\ 
\label{mnf3} F^{(3)} &=& 2 \, \alpha' g_s N \, \prod_{a=1}^3 
\left( \sigma^a - A^a \right) - \alpha' g_s N \, \sum_{a=1}^3 F^a 
\wedge \sigma^a 
\end{eqnarray} 
where 
\begin{eqnarray} 
\label{mnsol2} A^1 &=& - \frac{1}{2} \,a(\rho) \,d\theta_1 \spa A^2 = 
\frac{1}{2} \,a(\rho) \,\sin\theta_1 \,d \phi_1 \spa A^3 = - 
\frac{1}{2} \,\cos \theta_1 \,d\phi_1 \\ 
\label{mnsol3} e^{2h} &=&  \rho \coth 2 \rho - 
\frac{\rho^2}{\sinh^2 2 \rho} -\frac{1}{4} \spa a(\rho) = \frac{2 
\rho}{\sinh 2 \rho} 
\end{eqnarray} 
$A^a$ being the three $SU(2)_L$ gauge fields of the relevant 
seven-dimensional gauged supergravity. The $\sigma^a$ are the 
left-invariant one-form parameterizing the 3-sphere 
\begin{eqnarray} 
\label{sigmas} 
\sigma^1 &=& \frac{1}{2} \left(\cos \psi \,d\theta_2 + \sin \psi 
\sin\theta_2 d\phi_2\right) \spa  \sigma^2 = - \frac{1}{2} \left(\sin 
\psi \,d\theta_2 - \cos \psi \sin\theta_2 d\phi_2\right) \nonumber \\ 
\sigma^3 &=& \frac{1}{2} \left(d\psi + \cos\theta_2 d\phi_2 \right) 
\end{eqnarray} 

\section{ ${\cal N}=1$ super Yang-Mills theories in 4 dimensions}

We now show how to use the supergravity solution described in the previous 
paragraph to extract information on the corresponding
gauge theory. To this aim, we consider the world-volume action
 of a wrapped five-brane with a gauge field
strength $F$ and then extract from it the quadratic terms in $F$.
The non abelian bosonic action that we obtain is then simply \cite{DiVecchia:2002ks}:

\begin{equation}
S_{YM} = -\,\frac{1}{4g^{2}_{\rm YM}} \int d^4 x\,
 F_{\alpha \beta}^A \,F^{\alpha \beta}_A  + \frac{\theta_{\rm
 YM}}{32 \pi^2} \int d^4 x \,F_{\alpha \beta}^A\, {\widetilde{F}}^{\alpha
 \beta}_A \label{bound99}
\end{equation}
where
\begin{eqnarray}
\frac{1}{g^2_{\rm YM}}&=&\frac{\tau_5\,(2\pi)^2\alpha'^2}{2}
\int_{S^2}~ {\rm e}^{-\Phi}\,\sqrt{-\det G}~~,
\label{gym0} \\ \theta_{\rm YM} &=&{\tau_5\,(2\pi)^4\alpha'^2}
\int_{S^2}~ C^{(2)}~~.
\label{theta1}
\end{eqnarray}
where $C^{(2)}$ is the RR two form potential.

In the above set of coordinates, ($\theta_1,~\Phi_1,~\theta_2,~\Phi_2,~\psi$), 
the actual $S^2$ that enters these gauge-gravity relations is \cite{BM}
\begin{equation} 
\label{ciclo0} 
S^2 \ :\ \  \theta_1=-\theta_2 \spa \phi_1=-\phi_2 \spa \psi=0
\end{equation}

Note that the inverse square of the YM coupling constant is proportional to the volume
of the 2-sphere on which the D5 branes are wrapped and the YM vacuum
angle is proportional
to the flux of the R-R 2-form across this 2-sphere.

If we insert the explicit form of the supergravity
solution (\ref{mnsol})-(\ref{mnsol3})
in eq.(\ref{gym0}-\ref{theta1}), after simple calculations
we obtain
 
\begin{equation} 
\label{gym1} \frac{1}{g_{\rm YM}^2} = 
\frac{N}{16 \pi^2} \,Y(\rho) \ \ \ \ \ \ \ \ \ \ \ \ \ \  \ \ \theta_{\rm YM}=
-N\psi_0
\end{equation} 
where 
\begin{equation} 
Y(\rho) = 4 \,e^{2 h(\rho)} + \left( a(\rho) -1\right)^2 = 4 \rho 
\,\tanh \rho \spa a(\rho) = \frac{2\rho }{\sinh 2\rho}, 
\end{equation} 
$\psi_0$ being an integration constant. 

In the following we will exploit these results to discuss some
relevant features of the pure ${\cal N}=1$ SYM theory from the gravitational
point of view.

\subsection{The chiral symmetry}

As it is well-known, the ${\cal N}=1$ SYM theory possesses a
classical abelian $U(1)_R$ symmetry which becomes anomalous at the
quantum level. If we assign $R$-charge 1 to the gaugino
$\lambda(x)$ so that under a chiral transformation with parameter
$\varepsilon$
\begin{equation}
\lambda(x)~\to~{\rm e}^{{\rm i}\,\varepsilon}\,\lambda(x)~~,
\label{gaugin}
\end{equation}
then the presence of the anomaly implies that
\begin{equation}
\theta_{\rm YM} ~\to~\theta_{\rm YM} - 2 N\varepsilon~~.
\label{anom1}
\end{equation}

From eq.(\ref{gym1}) it is easy to see that we are describing the 
gauge theory in a fixed $\theta$ vacuum. Following a suggestion made in 
\cite{Maldacena:2000yy} we 
now argue that the other solutions are related to this by a seven dimensional
gauge transformation, namely
\begin{equation}
A'~=~ {\rm e}^{{\rm i}\varepsilon\sigma_3}~A~{\rm e}^{-{\rm i}\varepsilon\sigma_3}\label{tgauge}
\end{equation}

As pointed out in \cite{BM}, different seven dimensional gauge choices 
correspond to different parametrizations of the relevant ten dimensional 
geometry. In the case at hand, after the gauge transformation (\ref{tgauge}),
the longitudinal $S^2$ which enters the various gauge-gravity relations is now 
\begin{equation} 
\label{ciclot} 
S^2 \ :\ \  \theta_1=-\theta_2 \spa \phi_1=-\phi_2 \spa \psi= 2 \varepsilon
\end{equation}
From (\ref{gym0}-\ref{theta1}) it is easy to see that, on the gauge theory side, we are now describing the same running of the 
gauge coupling (as in (\ref{gym1})) but we are studying the theory in 
a different $\theta$-vacuum, namely:
\begin{equation}
\theta'_{{\rm YM}}=-N(\psi_0+ 2\, \varepsilon)
\end{equation}
The new form of the solution seems to suggest that even the phase of 
the gaugino condensate (that we identify with the supergravity field 
$a(\rho)$, see next section and \cite{Apreda:2001qb,DiVecchia:2002ks}) 
has been shifted by 
$2 \varepsilon$. We are then describing the different vacua related by the 
$U(1)_R$ symmetry of the gauge theory.

The $U(1)_R$ transformation which relates the different vacua is 
\begin{equation}
\psi~\to~\psi+\, 2\, \varepsilon~~, \label{psiepsi}
\end{equation}
This transformation is {\it not} a symmetry of the
supergravity solution (\ref{mnsol})-(\ref{mnf3}) (or of one related to
this by the gauge transformation (\ref{tgauge})) and is not
even an isometry of the metric. However, there is a region where 
the transformation (\ref{psiepsi}) is an isometry, 
namely the large-$\rho$ region
where the function $a(\rho)$ becomes exponentially small and can
be neglected. In fact, if we remove $a$, then all explicit $\psi$
dependence disappears from the metric (\ref{mnsol}) and the parametrization
of the world-volume 2-cycle no longer fixes the value of $\psi$ (differently 
from (\ref{ciclo0}-\ref{ciclot}) $\psi$ can be now 
an arbitrary constant).
The $\psi$ dependence instead still
remains in the R-R 2-form (\ref{mnf3}), which therefore is not
invariant under (\ref{psiepsi}). However the relevant quantity that 
should be considered is
the flux of $C^{(2)}$ across the 2-sphere, {\it i.e.}
\begin{equation}
\frac{1}{4\pi^2\alpha'g_s}\int_{S_2} \!\left.C^{(2)}\right|_{a=0}
=-\,\frac{N}{2\pi}(\psi-\psi_0)\label{flux1}
\end{equation}
which is allowed to change by integer values. Thus the
transformations (\ref{psiepsi}) with $\varepsilon=(\pi/N)k$ and $k$
integer are true symmetries of the supergravity background in the 
large-$\rho$ region. These are precisely 
the non-anomalous ${\mathbf Z}_{2N}$
transformations of the gauge theory that correspond to shifts of the
$\theta$-angle by integer multiples of $2\pi$. What we have described
here is therefore the gravitational counterpart of the well-known breaking
of $U(1)_R$ down to ${\mathbf Z}_{2N}$.

However, in the true supergravity solution $a(\rho)$ is not
vanishing and thus even the non-anomalous ${\mathbf Z}_{2N}$
transformations are not symmetries of the background. Thus,
the only non-anomalous symmetries of the supergravity background
are given by (\ref{psiepsi}) with $\varepsilon=k\pi$ and $k$
integer. This phenomenon is the gravitational counterpart of the
spontaneous breaking of the chiral symmetry from ${\mathbf Z}_{2N}
~\to~{\mathbf Z}_2$.

\subsection{The gaugino condensate}

In the previous subsection we have seen that the presence of a
non-vanishing $a(\rho)$ in the supergravity solution is
responsible for the spontaneous chiral symmetry breaking to
${\mathbf Z}_2$, which, on the gauge theory side, is accompanied
by the presence of a non-vanishing value of the gaugino condensate
$\langle \lambda^2 \rangle\equiv\langle 0|\left(\frac{{\rm
Tr}\,\lambda^2}{16\pi^2}\right)|0\rangle$. Thus, it appears very
natural to conjecture that the gravitational dual of this
condensate is precisely the function $a(\rho)$ that is present in
the supergravity solution (\ref{mnsol})-(\ref{mnf3}) 
\cite{Apreda:2001qb,DiVecchia:2002ks}

The gaugino condensate $\langle \lambda^2\rangle$ belongs to a
class of gauge invariant operators which do not acquire any
anomalous dimensions. This happens because the gaugino condensate
is the lowest component of the so-called anomaly multiplet whose
scale dimensions are protected by virtue of the fact that its top
component is the trace of the energy-momentum tensor which is a
conserved current. Thus, since the engineering dimension of
$\langle \lambda^2 \rangle$ is 3, we have
\begin{equation}
\langle \lambda^2 \rangle = c\,\Lambda^3 \label{condens}
\end{equation}
where $\Lambda$ is the (exact) dynamical scale of the ${\cal N}=1$
SYM theory and $c$ a computable constant.

In view of our previous
discussion, we now propose to identify the function $a(\rho)$ given in
(\ref{mnsol3}) with the gaugino condensate $\langle \lambda^2
\rangle$ measured in units of the (arbitrary) mass scale $\mu$
that is introduced to regulate the theory. Thus, we write 
\cite{DiVecchia:2002ks}
\begin{equation}
a(\rho) = \frac{\Lambda^3}{\mu^3}~~. \label{holo1}
\end{equation}
This crucial equation allows us to establish a precise relation
between the supergravity radial coordinate $\rho$ and the scales
of the gauge theory.
Notice that, as usual, since $a(\rho)\to 0$ for $\rho\to\infty$,
the large-$\rho$ region corresponds to the UV regime of the gauge
theory, and vice-versa the small-$\rho$ region corresponds to the
IR regime. 

Now we exploit the relation (\ref{holo1}) to
compute the perturbative $\beta$-function of the pure
${\cal{N}}=1$ SYM theory.

\subsection{The beta function}

From the above equations one can get the complete perturbative ${\cal 
N}=1$ $\beta$-function. We can write 
\begin{equation} 
\beta(g_{\rm YM}) = \frac{\partial g_{\rm YM}}{\partial \ln 
(\mu/\Lambda)} = \frac{\partial g_{\rm YM}}{\partial 
\rho}\frac{\partial \rho}{\partial \ln (\mu/\Lambda)} 
\end{equation} 
and compute the two derivative contributions from eq.~\eqref{gym1} and 
eq.~\eqref{holo1}, respectively. In doing so, let us first disregard 
the exponential corrections, which are sub-leading at large $\rho$ and 
which give rise to non-perturbative contributions. We easily get 
\begin{equation} 
\label{rolog1} \frac{\partial g_{\rm YM}}{\partial \rho} = - 
\frac{N g_{\rm YM}^3}{ 8 \pi^2} \spa \frac{\partial \rho}{\partial \ln 
(\mu/\Lambda)} = \frac{3}{2} \, \left(1 - \frac{1}{2 \rho}\right)^{-1} 
= \frac{3}{2} \, \left(1 - \frac{N g_{\rm YM}^2}{8 \pi^2}\right)^{-1} 
\end{equation} 
The final result is then 
\begin{equation} 
\label{betapert1} 
\beta(g_{\rm YM}) = - 3 \, \frac{N g_{\rm YM}^3}{16 \pi^2}  \left(1 - 
\frac{N g_{\rm YM}^2}{8 \pi^2} \right)^{-1} 
\end{equation} 
which is the NSVZ $\beta$-function \cite{Novikov:1983uc}.

Notice that the complete supergravity solution suggests the presence of a Landau pole. However, this is 
not really an issue. The curvature of the MN background goes like 
$\alpha'{\cal R} \sim 1/g_s N$ so the regime in which the supergravity 
approximation is reliable is for large $N$. In this regime a Landau 
pole can indeed be present even if the gauge coupling remains finite 
at the scale $\Lambda$, since in eq.~\eqref{gym1} it is really $g_{\rm 
YM}^2 N$ which is going to infinity for $\rho\to 0$ ($~ \mu\to\Lambda$) 
and not the gauge coupling 
itself. To discuss the duality in the deep IR at finite $N$, one has 
to go beyond the supergravity approximation. 


 
 


\section{Conclusions and Perspectives}

We have shown that many interesting features of the pure ${\cal
N}=1$ SYM theories in four dimensions are
quantitatively encoded in the supergravity solutions that describe
D5 branes wrapped on supersymmetric 2-cycles. These features
comprise the running of the gauge coupling constant, the
$\beta$-function and the chiral anomaly. 

All these are well known results on the gauge theory side, but the important 
point is that they are nicely encoded in a classical supergravity solution. 
Moreover, I find particularly nice that the gauginos condensation (a typical 
infrared phenomenum of the gauge theory) is also described by this 
supergravity solution. 

Another issue related to the infrared of the gauge 
theory is that of fractional instantons. The supergravity solution
 seems to suggest that they modify the
running of the Yang-Mills coupling constants itself \cite{DiVecchia:2002ks}. 
However in the 
IR these fractional instantons mix with some other unwanted Kaluza-Klein 
effects on the 
gravity side. This renders the correspondence with the gauge theory 
difficult to establish. Nevertheless one could infer that fractional instantons
with topological charge $2/N$ modify the running of the
coupling constant, giving a modified $\beta$-function with respect
to the NSVZ one. This would avoid the pole of the NSVZ $\beta$-function, whose
physical interpretation is not complety clear, giving:
\begin{equation} 
\label{betapert11} 
\beta(g_{\rm YM}) = - 3 \, \frac{N g_{\rm YM}^3}{16 \pi^2}  \left(\coth 
\Big( \frac{8\pi^2}{N g_{\rm YM}^2} \Big) - 
\frac{N g_{\rm YM}^2}{8 \pi^2} \right)^{-1} 
\end{equation} 

It is surely interesting to look for a precise form of the correspondence 
also in the infrared of the gauge theory.

\vspace{1 cm}

{\bf Acknowledgements}: 
I thank M. Bertolini, P. Di Vecchia, A. Lerda and W. Mueck 
for usefull discussions and/or correspondence.\\
I would like also to thank the organizers of the workshop 
``The quantum structure of spacetime and the geometric nature of fundamental
interactions'' in Leuven for the stimulating and nice atmosphere I have found 
there.\\
Work supported by the European Community's Human Potential
Programme under contract HPRN-CT-2000-00131 Quantum Spacetime
in which P.M. is associated to Copenhagen.


\small

\begin{thebibliography}{99}

\bibitem{Maldacena:2000yy} J. Maldacena and C. Nu\~nez, \emph{Toward 
The Large N Limit Of N=1 Super Yang Mills}, Phys. Rev. Lett. {\bf 86} 
(2001) 588, {\tt hep-th/0008001}. 
 
\bibitem{BM}
M.~Bertolini and P.~Merlatti,
arXiv:hep-th/0211142.

\bibitem{Vafa:2000wi} C. Vafa, \emph{Superstrings and topological 
strings at large N}, J. Math. Phys. {\bf 42} (2001), 2798, {\tt 
hep-th/0008142}. 
 
\bibitem{DiVecchia:2002ks} P.~Di Vecchia, A.~Lerda and P.~Merlatti, 
\emph{N = 1 and N = 2 super Yang-Mills theories from wrapped branes}, 
{\tt hep-th/0205204}. 
 
\bibitem{Loewy:2001pq} A. Loewy and J. Sonnenschein, \emph{On The 
Holographic Duals Of ${\cal{N}}=1$ Gauge Theories}, JHEP {\bf 0108} 
(2001) 007, {\tt hep-th/0103163}. 
 
\bibitem{Apreda:2001qb} R. Apreda, F. Bigazzi, A.L. Cotrone, 
M. Petrini, A. Zaffaroni, \emph{Some comments on N=1 gauge theories 
from wrapped branes}, Phys. Lett. {\bf B536} (2002) 161, {\tt 
hep-th/0112236}. 
 
\bibitem{Olesen:2002nh} P.~Olesen and F.~Sannino, \emph{N = 1 super 
Yang-Mills from supergravity: The UV-IR connection}, {\tt 
hep-th/0207039}. 

\bibitem{Cvetic:2000dm} M. Cvetic, H. Lu and C.N. Pope, 
\emph{Consistent Kaluza-Klein sphere reductions}, Phys. Rev. {\bf D62} 
(2000) 064028, {\tt hep-th/0003286}. 

 \bibitem{Chamseddine:1999uy} A.H. Chamseddine and W.A. Sabra, \emph{D 
= 7 SU(2) gauged supergravity from D = 10 supergravity}, 
Phys. Lett. {\bf B476} (2000) 415, {\tt hep-th/9911180}. 

\bibitem{Novikov:1983uc} V. Novikov, M. Shifman, A. Vainstein and 
V. Zakharov, \emph{Exact Gell-Mann-Low Function Of Supersymmetric 
Yang-Mills Theories From Instanton Calculus}, Nucl. Phys. {\bf B229} 
(1983) 381. 
 
\bibitem{cv} A.H. Chamseddine and M.S. Volkov, \emph{Non-abelian 
solitons in N=4 gauge supergravity}, Phys. Rev. Lett. {\bf 79} (1997) 
3343, {\tt hep-th/9707176}. 
 

\end{thebibliography}
\end{document}